# In-situ monitoring additive manufacturing process with AI edge computing

Wenkang Zhu, Hui Li, Yikai Zhang, Yuqing Hou, Liwei Chen

***Abstract*—In-situ monitoring system can be used to monitor the quality of additive manufacturing (AM) processes. In the case of digital image correlation (DIC) based in-situ monitoring systems, high-speed cameras were used to capture images of high resolutions. This paper proposed a novel in-situ monitoring system to accelerate the process of digital images using artificial intelligence (AI) edge computing board. It built a visual transformer based video super resolution (ViTSR) network to reconstruct high resolution (HR) videos frames. Fully convolutional network (FCN) was used to simultaneously extract the geometric characteristics of molten pool and plasma arc during the AM processes. Compared with 6 state-of-the-art super resolution methods, ViTSR ranks first in terms of peak signal to noise ratio (PSNR). The PSNR of ViTSR for 4× super resolution reached 38.16 dB on test data with input size of 75 pixels × 75 pixels. Inference time of ViTSR and FCN was optimized to 50.97 ms and 67.86 ms on AI edge board after operator fusion and model pruning. The total inference time of the proposed system was 118.83 ms, which meets the requirement of real-time quality monitoring with low cost in-situ monitoring equipment during AM processes. The proposed system achieved an accuracy of 96.34% on the multi-objects extraction task and can be applied to different AM processes.**

## I. Introduction

Additive manufacturing (AM) is the process of creating a part by joining material, typically by adding news layers over a substrate, in order to obtain a final product from data in a computer-aided design model [1]. The AM technology has gained tremendous interest from industry and academia because of its potential to manufacture complex components in a single stage [2]. However, it has limited acceptance in industries owing to its quality uncertainty including microstructural defects and residual stresses. Plasma arc additive manufacturing (PAM) process, a significant part of AM, is highly valued because of its potential for large-scaling manufacturing. Nevertheless, it is associated with dimensional inaccuracies and defects that hinder its further application in terms of high quality assurance. Meanwhile, the in-situ monitoring system is used widely for monitoring the quality of AM. To extract the features related to the product quality during manufacturing, Yang et al. [3] captured the shapes of spatters using max entropy method and revealed the relationship with number of spatters and laser power. Fang et al. [4] used a U-Net-based convolutional neural network (CNN) with lightweight architecture to accurately extract the molten pool signature. Similarly, Tan et al. [5] built a novel image segmentation network for spatter extraction involved with CNN-based selection and thresholding. Mi et al. [6] proposed a deep CNN to extract the geometric shape of molten pool and spatters simultaneously as molten pool and spatters are intra-related. Zhang et al. [7] used a fully convolutional network (FCN) to extract geometric properties of molten pool and plasma arc simultaneously in PAM. Up to 89% of full-field optical measurements use digital image correlation (DIC) [8], which relies heavily on the quality of images shot by the high-speed camera during the AM process.

With respect to the monitoring of the AM process, several image preprocessing methods are used to improve the image quality with the development of DIC. Luck et al. [9] got a series of normalized images by applying distortion correction to raw images using fully-constrained Homography matrix and light levelization process. Zhan et al. [10] constructed a multi-step image preprocessing method including graying, stretching method, histogram equalization, binarization, and morphological filtering to enhance the visual appearance of images. Scime and Beuth's work [11] used a Homography matrix and baseline intensity mask generated from an anomaly-free powder bed image to correct the distortion and remedy uneven lighting conditions. The above-mentioned image preprocessing methods can enhance the quality of raw images, but the fixed resolution of images limits the achievement of more accurate results.

In this paper, a novel in-situ monitoring system is proposed to extract the geometric characteristics of plasma arc and molten pool in PAM process from low resolution to high resolution. This system adopts the efficient and low-cost AI edge computing board as its computing center, instead of the traditional computer workstation. The high-resolution frames are reconstructed using a visual transformer based video super resolution model (ViTSR). The FCN takes the reconstruction of

This work was supported by the National Key Research and Development Program of China (No. 2022YFB4600800). Corresponding author: Hui Li.

Wenkang Zhu, Hui Li, Yikai Zhang, and Yuqing Hou are with the Institute of Technological Sciences, Wuhan University, 430072 Wuhan, China (e-mail: wenkang_zhu@whu.edu.cn; li_hui@whu.edu.cn, zhang_yikai@whu.edu.cn, houyuqing@whu.edu.cn).

Liwei Chen is with Department of Mechanical Engineering, Graduate School of Engineering, The University of Tokyo, 7-3-1 Hongo, Bunkyo-ku, Tokyo 113-8656, Japan (email: chen@hnl.t.u-tokyo.ac.jp).

ViTSR as input to simultaneously extract geometric characteristics of multiple objects. In Section 2, the details of the proposed system, such as including system architecture, AI edge computing board, video super resolution algorithm used in in-situ monitoring system, are described. In Section 3, the results of the in-situ monitoring system are shown; these results include the quality of 4× super resolution reconstruction and the extraction of geometric characteristics of molten pool and plasma arc during the AM processes. In Section 4, the performance of the proposed system is discussed. Finally, the conclusions are presented in Section 5.

## II. IN-SITU MONITORING SYSTEM WITH AI EDGE COMPUTING

### A. System architecture

Fig. 1 illustrates the architecture of the proposed in-situ monitoring system. As mentioned earlier, this system can extract the high-resolution geometric characteristics in PAM process. The system includes the powder feed PAM equipment (ABB IRB 2600, Guangzhou LeiJia Additive Manufacturing Technology Co., Ltd., China), AI edge computing board (Jetson Xavier NX, NVIDIA Co., Ltd., USA), and a high-speed camera with maximum sample rate of 30000 fps (MEMRECAM ACS-1, NAC Image Technology Inc., Japan).

The resolution and sample rate of the high-speed camera can be adjusted manually. The video sequence from the high-speed camera is first sent to the AI edge computing board via universal serial bus (USB) interface, and then the super resolution frames are exported to the monitor via high-definition multimedia interface (HDMI) interface. The video sequence is acquired and encoded by OpenCV [12]. Neural networks are programmed by Python 3.8 using TensorFlow 2.5 and then optimized by TensorRT to accelerate the inference process on the neural process unit (NPU) of AI edge computing board.

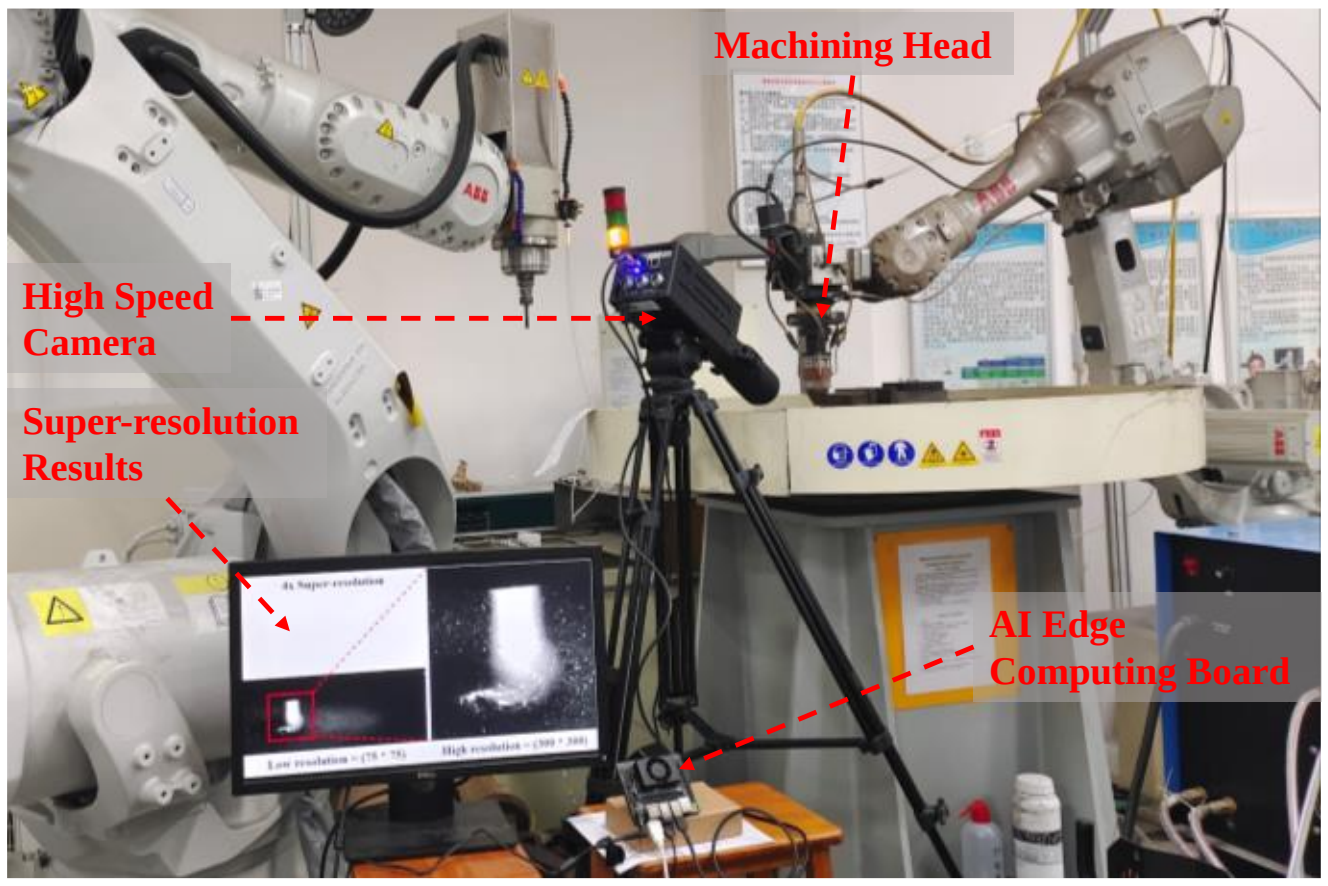

Fig. 1. In-situ monitoring system integrated with AI edge computing board during PAM process.

### B. AI edge computing board

This AI edge computing board uses Jetson Xavier NX as system-on-a-chip (Soc), which integrates CPU, GPU, and NPU into a single circuit by NVIDIA Co., Ltd., USA. Jetson Xavier NX is designed for high-performance and energy-efficient usage at a low cost. The technical specification of Jetson Xavier NX is given in Table I. It runs Jetson operating system based on Ubuntu 18.04 with JetPack software development kit (SDK). The system also contains Linux driver packages, CUDA libraries, and related application programming interfaces.

TABLE I
TECHNICAL SPECIFICATION OF JETSON XAVIER NX [13]

| Feature | Description |
|---|---|
| AI performance | 21 TOPS (INT8) |
| GPU | 384-core Volta$^{TM}$ GPU |
| CPU | 6-core Nvidia Cameral CPU ARM®v8.2 |
| RAM | 8 GB 128-bit LPDDR4x |
| Tensor cores | 48 |
| Power | 10 W \| 15 W \| 20 W |

Fig. 2 shows the overall layout and components of the AI edge computing board. This board contains two CMOS serial interfaces for interface. It also contains an HDMI, multiple USB 3.0 interfaces, and micro USB interface for display and data interaction. A wide area network port and a Bluetooth chip are equipped on the board for wireless transfer of data. The fixed 128 GB solid state disk meets the requirements of space for storing different AI models, SDK, and deep learning framework.

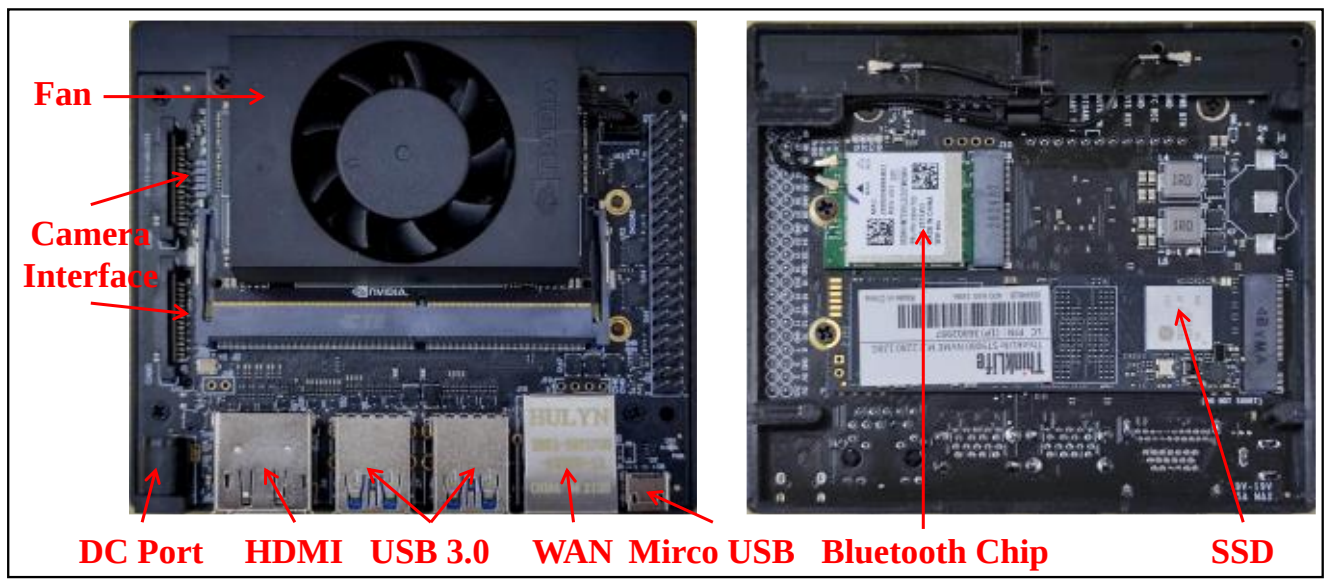

Fig. 2. AI edge computing board for video decoding and inference of AI models during in-situ monitoring of PAM.

### C. Video super resolution algorithm for low resolution video sequence

Traditional single image super resolution [14]-[16] algorithms often lead to blurry effects and motion artifacts for the cause of excessive destruction of ground truth textures and missing consideration of temporal relationship [17]. A previous study [18] used interaction-learning strategy to reduce computation, although the complex architecture is unsuitable to parallel acceleration. Most of the popular VSR methods [19], [20] adopt the fixed pipeline of motion estimation, motion compensation, fusion, and upsampling. The optical flows between frames are first estimated and then used to align features so as to eliminate the motion effects. However, these methods heavily rely on optical flow estimation, which is complex and time-consuming. Thus, more deep neural networks [21], [22] are proposed to compensate the motion by implicit estimation.

Super resolution methods are commonly used to enhance the quality of raw images during preprocessing. Shen et al. [23] applied smooth and sparse tensor completion in data preprocess of AM to propose a super resolution method for multi-sources image stream data. Walecki et al. [24] improved the confidence of object surface by using multiple images to tighten the line segments along the camera ray. However, these traditional mathematical-based methods are limited by complex task settings.

In this paper, a super resolution model based on a visual transformer is proposed to upscale the input. Fig. 3 shows the overall pipeline of ViTSR. The ViTSR considers a sequence of low resolution (LR) frames $\{X_{t-N},...,X_t,...X_{t+N}\}$ as its input, where $X_t$ represents the reference frame and the others stand for neighboring frames. The output of ViTSR is $Y_t$, which refers to the high resolution (HR) version of the reference frame $X_t$.

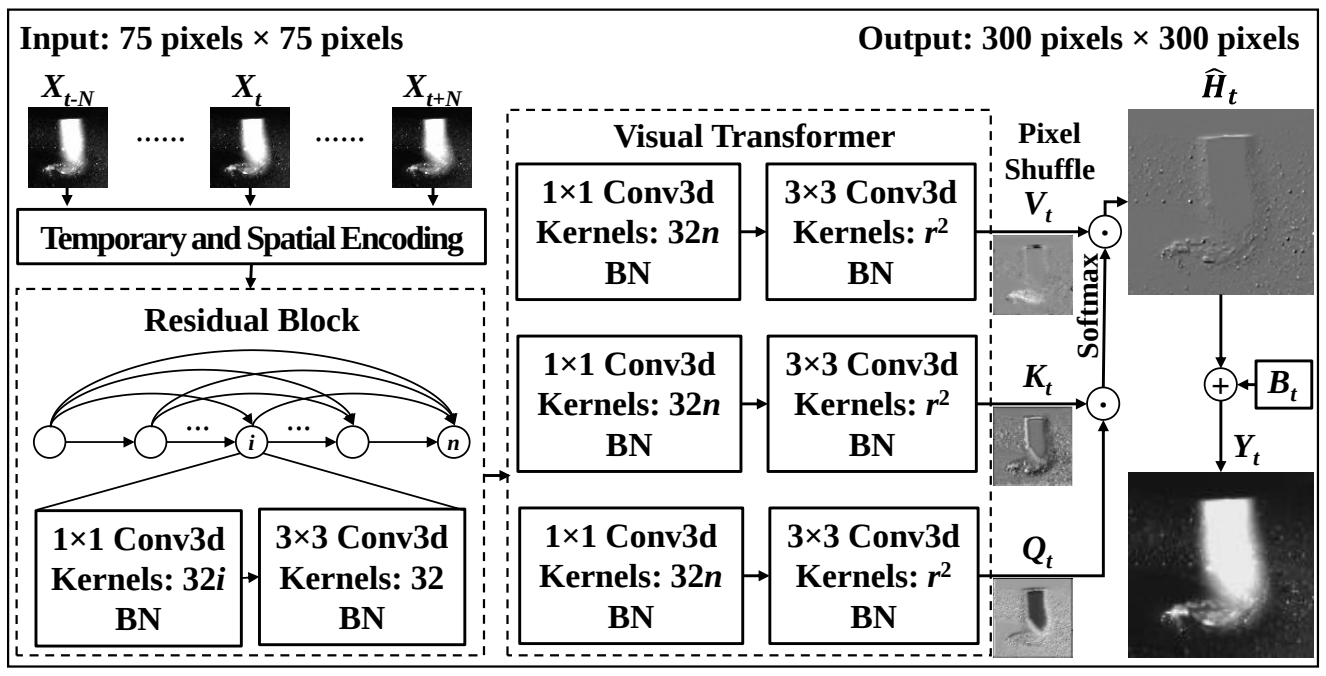

Fig. 3. Video super-resolution using visual-transformer.

LR frames are downscaled from the corresponding ground truth (GT) frames by nearest neighbor interpolation with scale of *r*. The height and width of single input frame are represented as *H* and *W*. As per the standard practice, RGB images are converted into Y-Cb-Cr color space, and only the Y channel is used for super resolution [25].

In addition, temporary and spatial information is explicitly encoded and concatenated to the input frames at first. Given a pixel with coordinates (*x*, *y*, *i*) in the frame $X_i$, where *x*, *y*, and *i* imply the horizontal, vertical, and temporal codes of this pixel, temporary, and spatial encoding of this pixel can be formulated as:

$$[h(x),v(y),t(i)] = [\sin(\frac{x}{W}-\frac{1}{2}),[\sin(\frac{y}{H}-\frac{1}{2}),[\sin(\frac{i}{T}-\frac{1}{2})] \quad (1)$$

where $h(x)$, $v(x)$, and $t(i)$ rapidly stand for horizontal code, vertical code, and temporal code of this pixel, respectively.

A residual block with *n* cells serves as a feature extractor, inspired by [26]. Each cell receives the concatenation of the internal outputs before itself and finally exports a feature map with 32 channels. As shown in Fig. 3, every single cell consists of convolutional layers with batch normalization. The residual cell *i* ($1 \le i \le n$) consists of 1 × 1 convolutional layers, as described previously [27], and 3 × 3 convolutional layers to enhance feature interaction.

After the residual block, a visual transformer block is followed. The visual transformer block is composed of 3-way parallel cells with an inner structure similar to residual cell. *Q* (Query), *K* (Key), and *V* (Value) indicate three essential tensors in the attention mechanism [28] that will be used to calculate the fused feature map. In particular, the last way in the block applies no padding in the temporary axis of feature map, which leads the shape of *Q* tensor to $1 \times H \times W \times r^2$. In addition, the *K* tensor includes all key features of neighboring frames with the shape of $2N \times H \times W \times r^2$, and *V* tensor includes all value features of frames with the shape of $(2N+1) \times H \times W \times r^2$. The fused feature map $H_t$ can be calculated as

$$H_t = V_t + \sum_{i=t-N, i\neq t}^{t+N} Softmax(Q_t, K_i)V_i \quad (2)$$

where $K_i$ and $V_i$ refer to the slices of *K* and *V* on the temporary axis, and $Q_t$ equals to *Q*. Then a pixel shuffle, which is a periodic shuffling operator [13], is applied to rearrange the tensor $H_t$ to $\hat{H}_t$. The reconstructed frame $Y_t$ is produced by the sum of bicubic interpolation $B_t$ and $\hat{H}_t$.

## III. Results

### A. Video sequence captured by in-situ monitoring system

During the PAM process, video sequences containing molten pool and plasma arc are captured by the high-speed camera at a current intensity of 40 A and scanning speed of 10 mm/s. Fig. 4 shows the original images captured by the high-speed camera with no preprocess in 1 ms. Frames in Fig. 4 show the temporal continuity of motion, indicating the correlation of neighboring frames that can be used to enhance the feature of the reference frame. Our findings showed that the plasma arc sways over time and molten pool flows slowly.

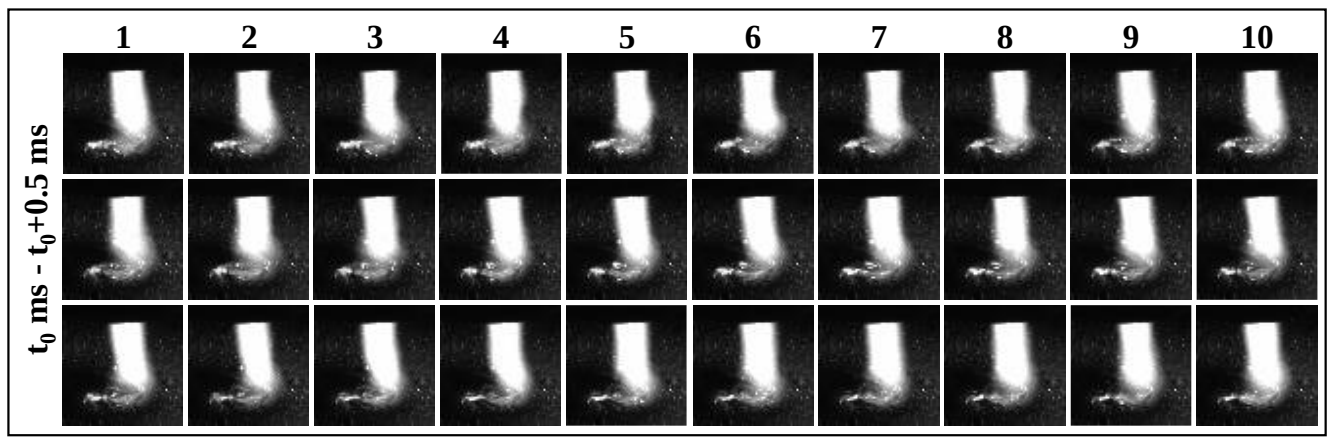

Fig. 4. Original video of molten pool and plasma arc captured at a current intensity of 40 A and scanning speed of 10 mm/s.

### B. In-situ video super resolution reconstruction of monitored videos

The proposed ViTSR was used to reconstruct the frames obtained from the high-speed camera to obtain high resolution video sequence. Fig. 5 shows the high-resolution reconstruction of local regions in video sequence by ViTSR, where *N* is set to 1 and upscaling ratio *r* is 4. The input of the super resolution model includes one reference frame and two neighbor frames. Here, $t_0$ is the start time of video sequence, and T is the interval time of shooting. In the figure, the images on the left show the position of the regions to be reconstructed, while the images on the right are coupled with low-resolution reference frames and corresponding reconstructed frames. The top part of images shows regions of powder bed, the middle part shows regions of plasma arc, and the bottom part shows molten pool. In the proposed VSR model, the resolution of the images is enlarged from 25 pixels × 25 pixels to 100 pixels × 100 pixels. The motion between high-resolution frames is natural and reasonable because the visual transformer enables implicit motion estimation and pixel-wide fusion.

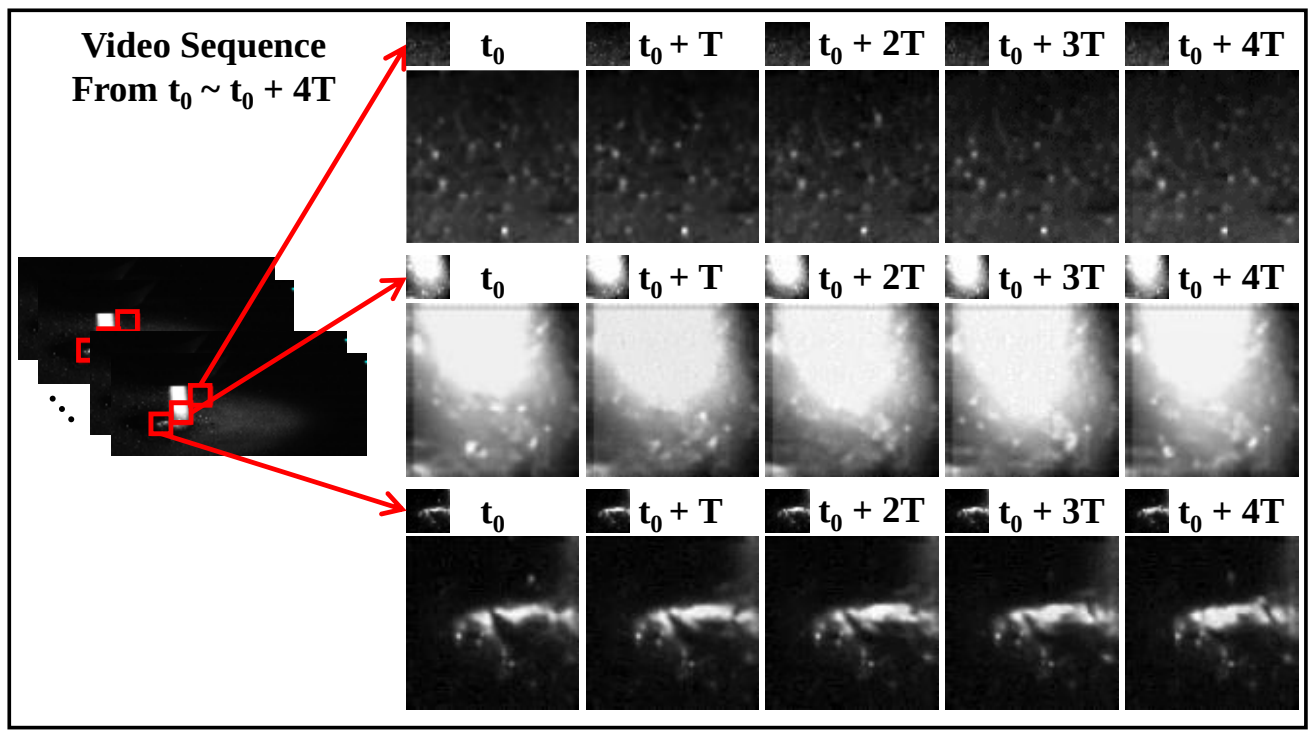

Fig. 5. Super-resolution for molten pool and spatter areas in a video sequence by ViTSR, T = 1/30 ms.

In Fig. 6, critical internal feature maps were considered for visualization to evaluate the visual transformer block with respect to feature fusion. The corresponding internal feature maps of input frames are $\{X_{i-1}, X_i, X_{i+1}\}$. Meanwhile, *Q*, *K*, and *V* are sliced as $Q_i$, $\{K_{i-T}, K_{i+T}\}$, and $\{V_{i-T}, V_i, V_{i+T}\}$ by the axis of channel, which is also the last channel of feature maps. In Fig. 6, the visualization of feature maps *Q*, *K*, and *V* have profiles similar to the corresponding input frames, including the topography of the plasma arc and molten pool, and the distribution of metal powder. The fused feature map $H_i$ shows details of the fusion of *Q*, *K*, and *V* on the way of attention mechanism, which keeps the regions that can be used for the reconstruction of the reference frame and suppresses those regions that are different from the reference frame. The visualization of $H_i$ shows great similarity of geometric shape with reference frame. This indicates that high attention is paid to similar regions of neighbor frames and reference frame, while low attention is paid to different regions, indicating the effectiveness of this block. When applying visual transformer block instead of motion estimation and feature alignment, the beneficial features from neighbor frames can be fused into the correct regions of reference frame by using parallel neural nodes and nonlinear combination.

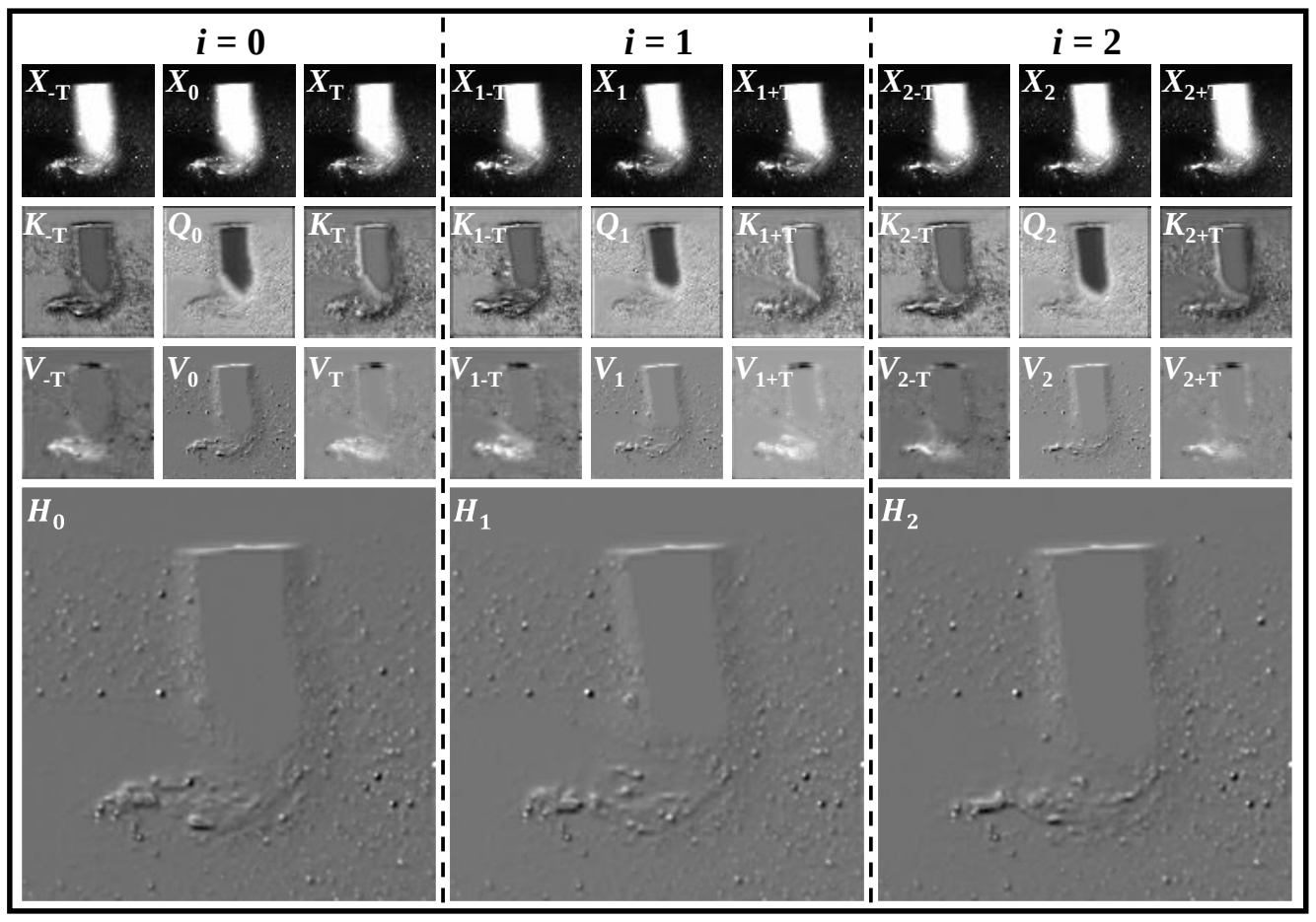

Fig. 6. The attention map *H* decoded from query *Q*, key *K* and value *V*.

The proposed model was compared with 6 state-of-the-art super resolution methods, including Bicubic, super resolution convolution neural network (SRCNN) [15], fast SRCNN (FSRCNN) [16], efficient sub-pixel convolutional neural network (ESPCN) [14], dynamic upsampling filters (DUF) [21], and temporary group attention (TGA) [22]. Bicubic is a simple math-based method that is widely used for its low cost. The SRCNN applied convolutional layers behind bicubic interpolation to seek higher quality of HR reconstruction. Deconvolution is used in FSRCNN to upscale the output instead of interpolating the input image in the beginning so as to make full use of feature maps. Pixel shuffle is used in ESPCN to obtain surprising results. DUF applies dynamic upsampling filters to compensate for motion between frames implicitly instead of traditional optic-flow motion estimation. The reconstruction in TGA can be improved using temporary group attention for the inner relationship between long-term frames. Bicubic interpolation realized by OpenCV is directly used for its high efficiency. Fig. 7 shows comparisons between 4× super resolution reconstruction by the state-of-the-arts models and the proposed model. The input size of models was set to 75 pixels × 75 pixels, which is LR in the figure. Three reference frames were selected with interval time ($T_S$) of 1/3 ms. Finally, ViTSR ranks first in terms of PSNR and was 0.7770 dB higher than the second highest model.

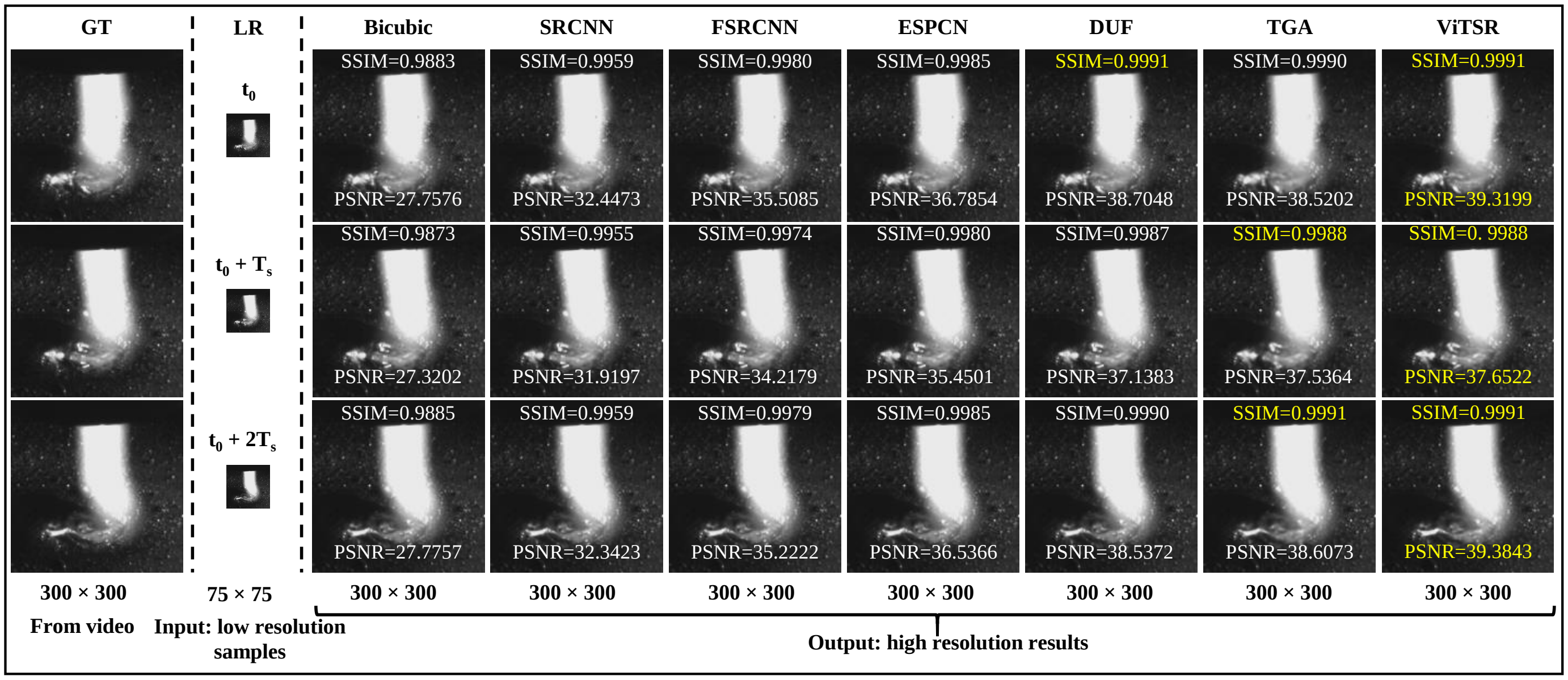

Fig. 7. Performance of super-resolution results by various methods.

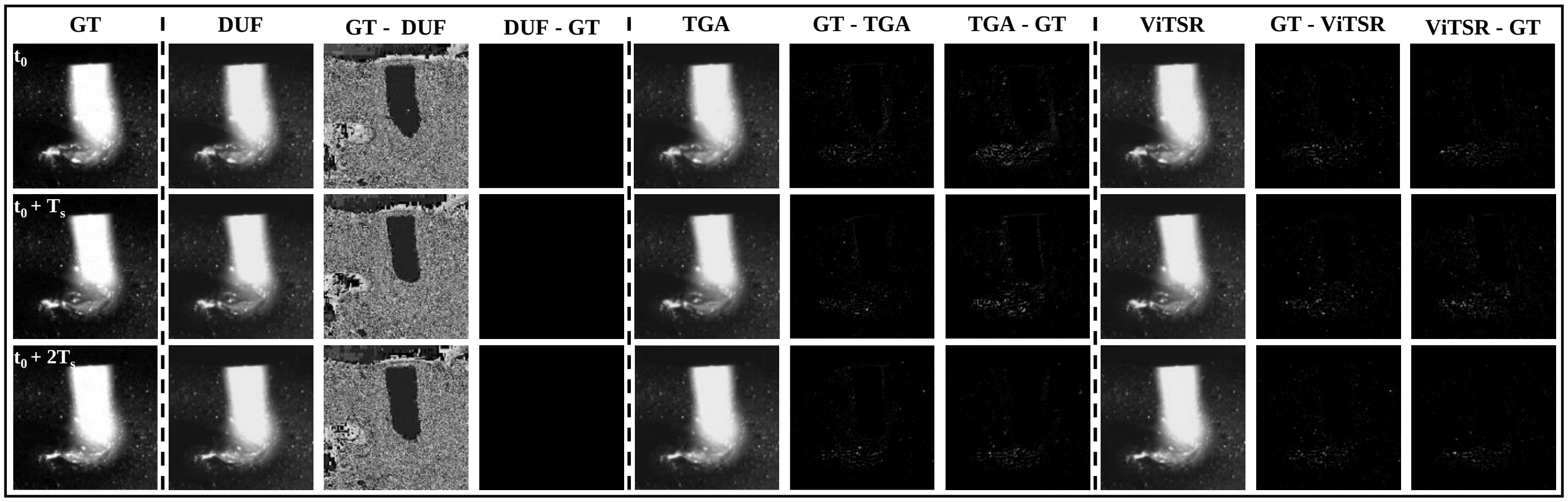

Fig. 8. Visualization of the difference between super-resolution reconstruction and GT, $T_s$ = 1/3 ms.

The top 3 models (ViTSR, TGA and DUF) were selected in Fig. 8 to visualize the subtracting of reconstructed frames with GT frames and to clearly show the difference between reconstructed frames and GT frames. Negative values of subtraction were set to zero, and the values of subtraction were scaled to 0–255. As shown in Fig. 8, most areas of DUF were different with GT frames for its dynamic filters, and the results of DUF showed lower brightness than GT frames. For TGA, the different pixels were concentrated around molten pool and plasma arc. In the case of ViTSR, the different pixels were concentrated around molten pool, and the number is the least.

Fig. 9 shows the average PSNR and inference time of various methods on the AI computing board with the same setting. A total of 58 video frames were tested. As shown in Fig. 9, ViTSR achieved the best performance in terms of PSNR but only half-time cost compared with the TGA. After time optimizing using operator fusion and model pruning, ViTSR performs the best by making most of the NPU. Inference time of ViTSR was 233.01 ms on CPU that was optimized to 50.97 ms on AI edge board. The PSNR of ViTSR was reduced to 37.67 dB, which indicates saved inference time.

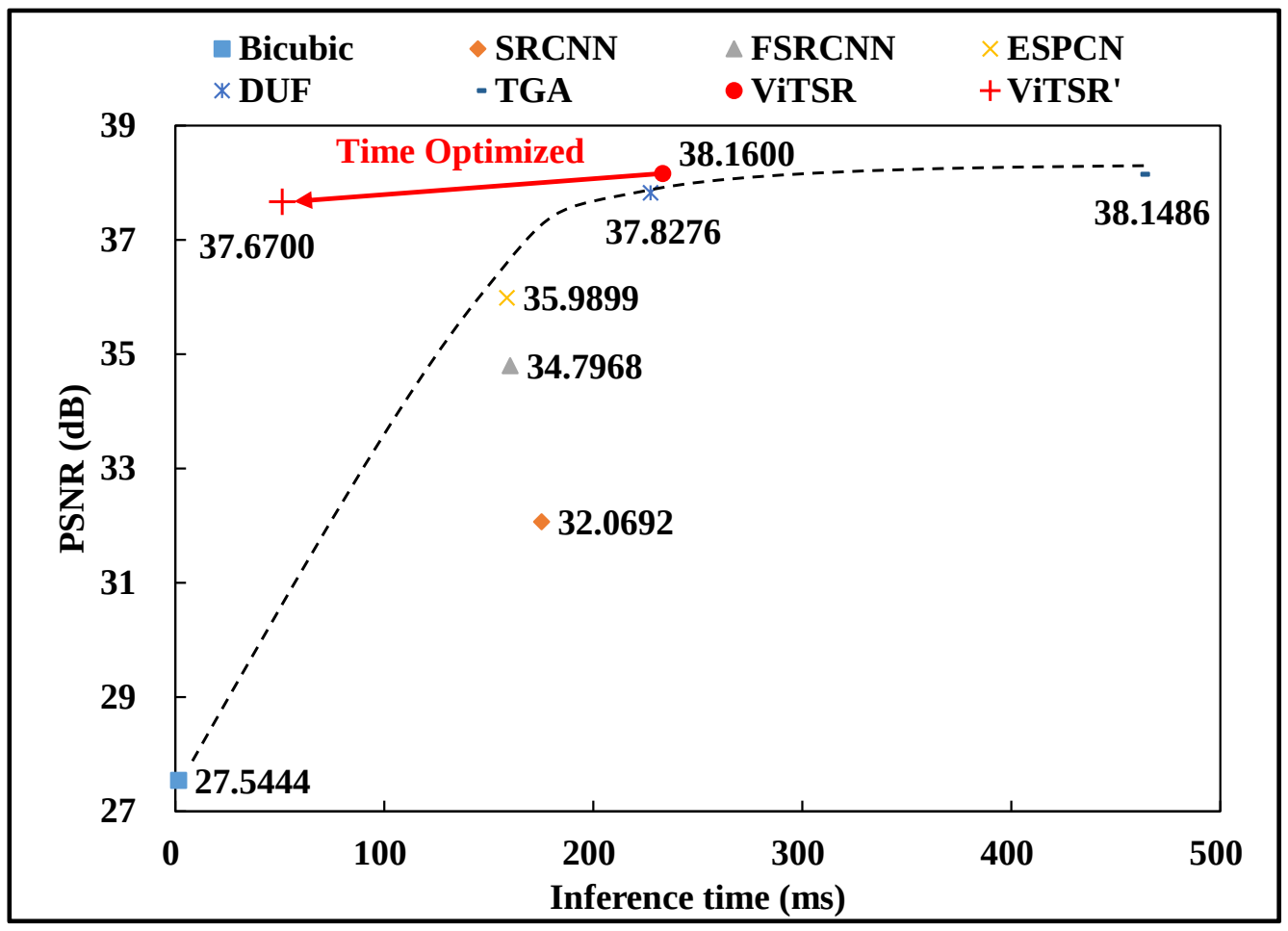

Fig. 9. Performance of various models in terms of PSNR and inference time on AI computing board.

The effectiveness of core components of this model, including the temporal and spatial encoding, visual transformer, and 3D convolution, was verified. The core components of ViTSR were eliminated to generate three degenerated models, named as ViTSR_d1, ViTSR_d2, and ViTSR_d3. In brief, ViTSR_d1 removed temporal and spatial encoding from the proposed model, ViTSR_d2 used DUF instead of the visual transformer, and ViTSR_d3 replaced all 3D convolution with 2D convolution. Table II shows the PSNR of ViTSR and degenerated models on the test data after training. The ViTSR showed the highest PSNR (38.16 dB) with 0.96 dB higher than the second-best model.

TABLE II
PSNR OF ViTSR AND DEGENERATED MODELS ON TEST DATA AFTER TRAINING

| Models | Description | PSNR |
|---|---|---|
| ViTSR | the proposed model | 38.16 dB |
| ViTSR_d1 | remove temporal and spatial encoding | 36.42 dB |
| ViTSR_d2 | replace VIT with DUF | 36.29 dB |
| ViTSR_d3 | replace 3D-Conv with 2D | 37.20 dB |

### *C. In-situ extraction of geometric characteristics of molten pool and plasma arc*

In our previous work [6], plasma arc and molten pool were simultaneously extracted in the process of PAM. The present paper used the well-trained FCN to extract the geometric characteristics of molten pool and plasma arc after ViTSR. Fig. 10 shows the segmentation of the proposed in-situ monitoring system. The resolution of input video sequence was 75 pixels × 75 pixels, while the output resolution of segmentation was 300 pixels × 300 pixels after ViTSR and FCN.

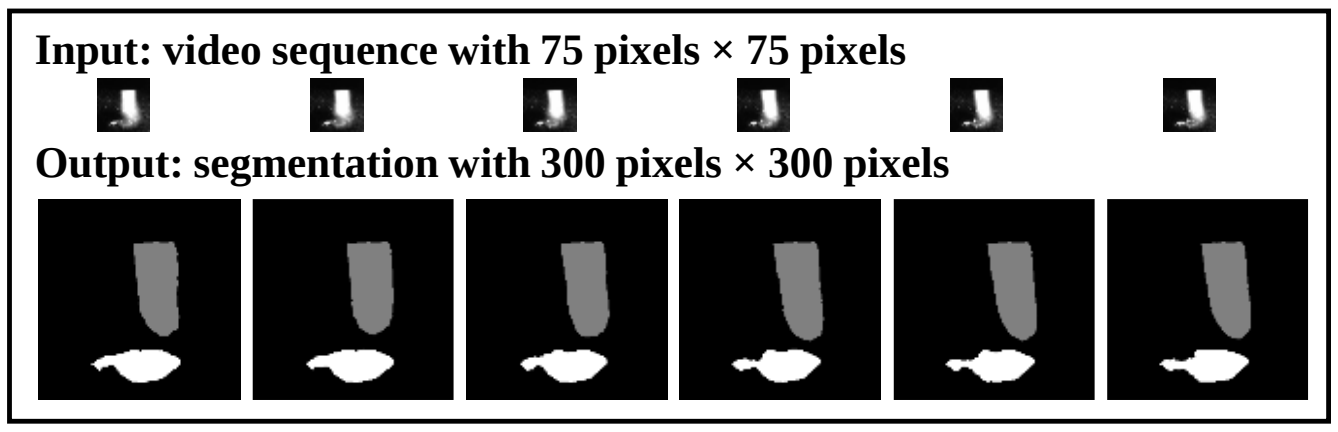

Fig. 10. Extraction of geometric characteristic of in-situ monitoring video.

The AI edge computing used ViTSR and FCN to fetch the areas of real-time super resolution segmentation. The ViTSR

was used to reconstruct HR frames, while the FCN was used to extract the geometric shapes of molten pool and plasma arc. Fig. 11 shows the extracted pixels of molten pool and plasma arc in PAM process with time. The area of molten pool and plasma arc was found to be in the range of 3383–4135 pixels and 7474–8709 pixels, respectively. This shows the temporal coherence of the target areas. As shown in this figure, the AI edge computing board was used to process 58 frames with a resolution of 75 pixels × 75 pixels to 300 pixels × 300 pixels results. The video sequence frames with a total number of 58 were processed for super resolution reconstruction and frame segmentation in 6.89 s. The average inference time for processing a single frame is 118.83 ms, where the inference time of frame segmentation is 67.86 ms. The in-situ monitoring system showed its immense future for high-resolution and low-cost quality monitoring of AM process.

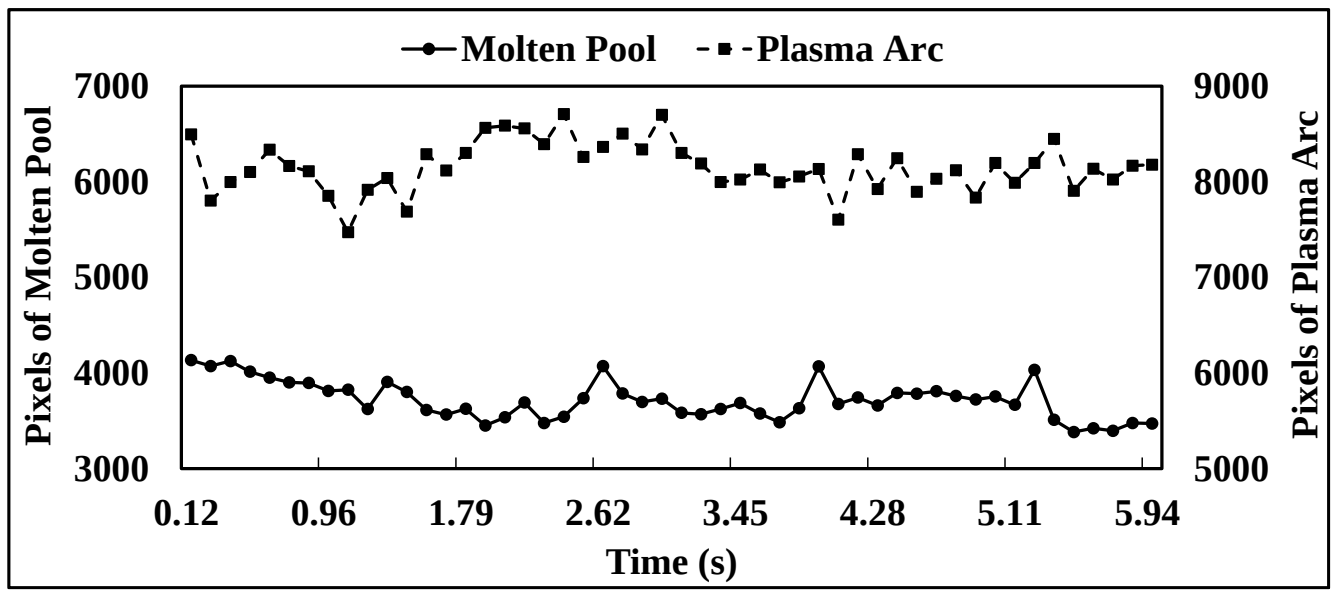

Fig. 11. Extracted pixels of molten pool and plasma arc with time.

## IV. Discussion

Traditional image segmentation algorithms were successfully applied in many fields, whereas the in-situ monitoring task of AM could not fit because of its halo effects and high dynamic range imaging. Fig. 12 shows the comparison of extraction performance using traditional and proposed methods with input size of 75 pixels × 75 pixels in PAM process. As shown in Fig. 12(a), plasma arc, molten pool, and part of powder bed were segmented together when using triangle algorithm. As shown in Fig. 12(b), the maximum entropy algorithm extracted the general shapes of plasma arc, molten pool, and bright metal powders. The region of plasma arc extracted by the two methods was severely enlarged for interference of halo. With respect to the extraction of targets and the suppression of bright metal powders, watershed in Fig. 12(c), Otsu in Fig. 12(d), and basic global thresholding algorithm in Fig. 12(e) performed better than the former segmentation algorithms. The three algorithms were not critically affected by the halo and reflected light, while the extraction of plasma arc was still enlarged and the molten pool was smaller than the fact. Traditional algorithms could not extract target areas accurately due to the interference of halo and reflection. As shown in Fig. 12(f), FCN extracted the plasma arc and molten pool efficiently owing to its nonlinear fitting ability, where the halo was excluded from the extraction and the low brightness showed less interference of the results. Fig. 12(g) showed the extracted result of 4× super resolution frame using ViTSR and FCN. Our findings showed that the proposed method extracted plasma arc and molten pool accurately, eliminating the interference of light conditions and bright powders. Table III shows the performance of traditional methods, FCN method, and the proposed method. The table also showed that the proposed method upscaled the resolution of input frames from 75 pixels × 75 pixels to 300 pixels × 300 pixels, thus, achieving the highest accuracy.

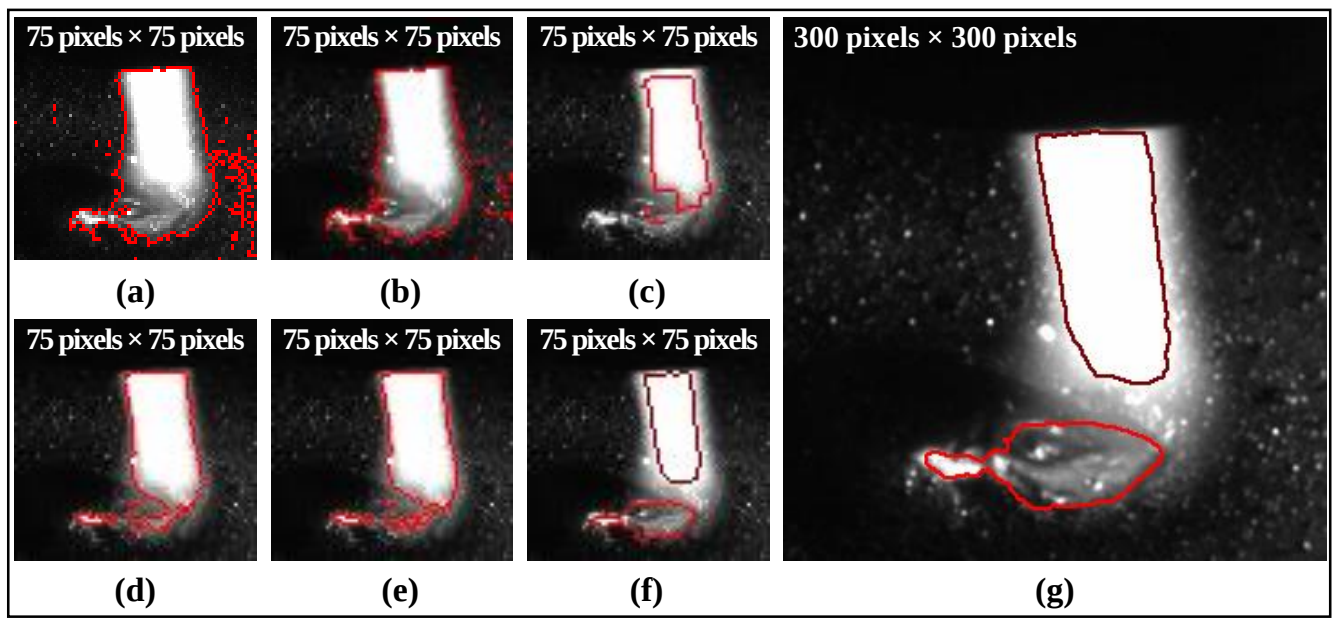

Fig. 12 Extraction of molten pool and plasma arc using different image segmentation methods: (a) triangle, (b) maximum entropy, (c) watershed, (d) Otsu, (e) basic global thresholding, (f) FCN, and (g) ViTSR+FCN.

TABLE III
PERFORMANCE OF TRADITIONAL SEGMENTATION METHODS AND PROPOSED METHOD

| Methods | Extracted objects | Resolution (pixel) | Upscaling | Accuracy |
|---|---|---|---|---|
| Triangle | Bright area & Dark area | 75 × 75 | 1 × 1 | 27.57% |
| Maximum Entropy | Bright area & Dark area | 75 × 75 | 1 × 1 | 85.46% |
| Watershed | Bright area & Dark area | 75 × 75 | 1 × 1 | 92.11% |
| Otsu | Bright area & Dark area | 75 × 75 | 1 × 1 | 90.57% |
| Basic Global Thresholding | Bright area & Dark area | 75 × 75 | 1 × 1 | 92.68% |
| FCN | Plasma arc & Molten pool | 75 × 75 | 1×1 | 96.29% |
| Proposed: ViTSR + FCN | Plasma arc & Molten pool | Input: 75 × 75 Output: 300 × 300 | 4×4 | 96.34% |

Table IV showed the performance of AI methods for the in-situ monitoring of multiple AM process. Tan's method [4] was used to process the image tiles with a resolution of 200 pixels × 200 pixels so as to detect the spatters in the process of laser power bed fusion (LPBF). Despite the acceptable inference time, the accuracy may not be suitable for real-world applications. Fang reported high accuracy and inference speed after he adopted U-Net to extract the molten pool with a resolution of 224 pixels × 224 pixels. Mi et al. [5] achieved an accuracy of 94.71% in the process of laser based directed energy deposition (L-DED) when the proposed D-CNN architecture was used to extract the spatters and molten pool simultaneously. Zhang et al. [6] extracted plasma arc and molten pool simultaneously in the PAM process. Although the inference time of Zhang's method [6] was longer than that of Mi's method [5], the accuracy achieved by using Zhang's method [6] was much higher. In this paper, the proposed method was used to reconstruct the 4× super resolution frames and extract the geometric characteristics of plasma arc and molten pool during PAM process. The proposed method achieved an accuracy of 96.34% with inference time of 15 ms and 118.83 ms on Nvidia RTX 3070 Laptop GPU and Nvidia Jetson Xavier NX, respectively. The sum of two models denoted the number of parameters; this is one of the reasons of the higher inference time. In the case of less power, the AI edge computing board was equipped with 48 tensor cores and only 26% of GPU, which is the reason of higher inference time. The inference time of the proposed method using GPU was found to be the smallest in Table IV even when the GPU had more parameters, which shows the efficiency of the proposed method. The inference time can be optimized in the future by integrating the FCN and ViTSR into one step. The proposed method shows its low requirement of pixel resolution for high-speed camera used in the video capture of the AM process. This reduced the cost of camera and contributed to the industrial applications of in-situ monitoring system for AM.

TABLE IV
PERFORMANCE OF OUR PREVIOUS METHODS AND THE PROPOSED METHOD IN THIS WORK FOR IN-SITU MONITORING OF PAM

| Methods | Extracted objects | Resolution (pixel) | Upscaling | Inference time Accuracy |
|---|---|---|---|---|
| Tan's [4] | Spatters | 200 × 200 | 1 × 1 | 70 ms 80.48% |
| Fang's [3] | Molten pool | 224 × 224 | 1 × 1 | 37 ms 98.06% |
| Mi's [5] | Spatters & Molten pool | 450 × 512 | 1 × 1 | 63 ms 94.71% |
| Zhang's [6] | Plasma arc & Molten pool | 450 × 512 | 1 × 1 | 84 ms 95.10% |
| Proposed: ViTSR + FCN | Plasma arc & Molten pool | Input: 75 × 75 Output: 300 × 300 | 4 × 4 | 15 ms (GPU) 118.83 ms (AI edge computing board) 96.34% |

## V. CONCLUSIONS

In this paper, a novel in-situ monitoring system was proposed to extract the high-resolution features of the AM process. For this, high-speed camera with variable resolutions was used to capture the video data during the AM processes. The captured video sequences were processed using AI edge computing because of its low cost and high efficiency. The maximum power of the AI edge computing board was found to be 20 W, which is lower than the power consumption of PCs and computing servers. The proposed system used 118.83 ms for total inference of video sequences. The inference time of video super resolution is 50.97 ms from a resolution of 75 pixels × 75 pixels to 300 pixels × 300 pixels, and frame segmentation is 67.86 ms, respectively. Two-stage strategy is used to reconstruct the input with high resolution and extracts the key features with high accuracy. To the best of our knowledge, this is the first study that used a video super resolution algorithm before image segmentation to seek high resolution geometric characteristics of AM processes. The output of the proposed system demonstrated greater tolerance to halo and shadow of captured video and finally achieved an accuracy of 96.34%, which is similar to our previous work. Considering that the proposed system needed 1/4 resolution as input with same results, the system will sharply reduce the cost of high-speed cameras used in in-situ monitoring systems. Thus, this paper provides a way to lower the cost of DIC-based methods and may finally improve the quality of products manufactured by various AM process including PAM, LPBF, and L-DED.

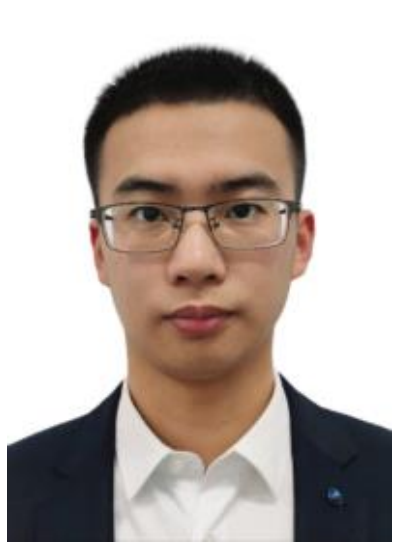

**Wenkang Zhu** received the B.S. degree in mechanical engineering and automation from Wuhan University, Wuhan, China in 2017.

He is studying for master's degree in Institute of Technological Sciences, Wuhan University, Wuhan, China. His research interests include additive manufacturing and artificial intelligence.

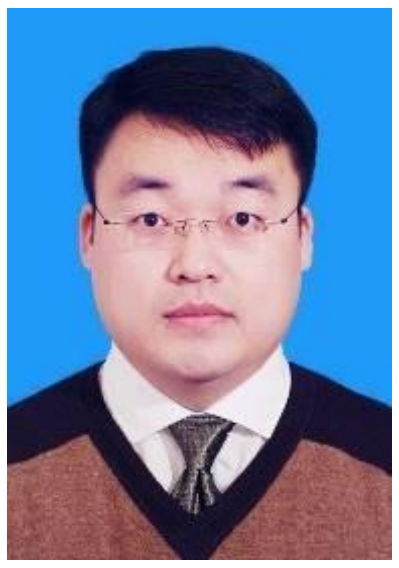

**Hui Li** received the B.S. degree from Huazhong University of Science and Technology, Wuhan, China, and the Ph.D. degree in electrical & computer engineering from National University of Singapore, Singapore, in 1999 and 2007, respectively.

He is currently a Professor at Wuhan University, China. His research interests include electronics manufacturing and additive manufacturing.

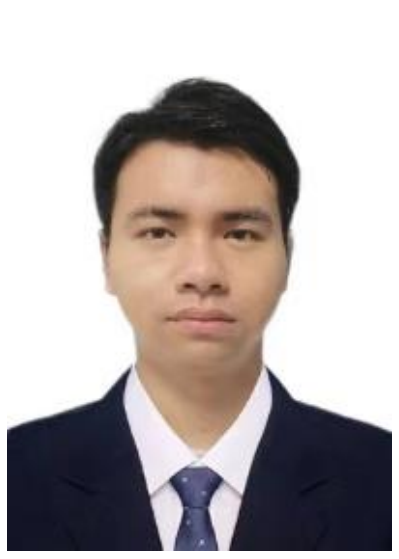

**Yikai Zhang** received his B.S. degree in electronic information engineering from Hefei University of Technology in 2020 and is currently studying for a master's degree at the Institute of Industrial Science of Wuhan University.

His current research interests include image processing, deep learning, and additive manufacturing.

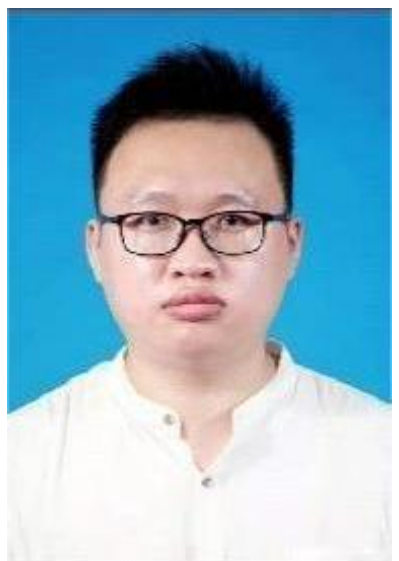

**Yuqing Hou** received his M.S. degree from Xian University of Technology, Xian, China. He is currently pursuing the Ph.D. degree with Institute of Technology Science, Wuhan University, Wuhan, China.

His research focuses on artificial intelligence and computational fluid dynamics for additive manufacturing.

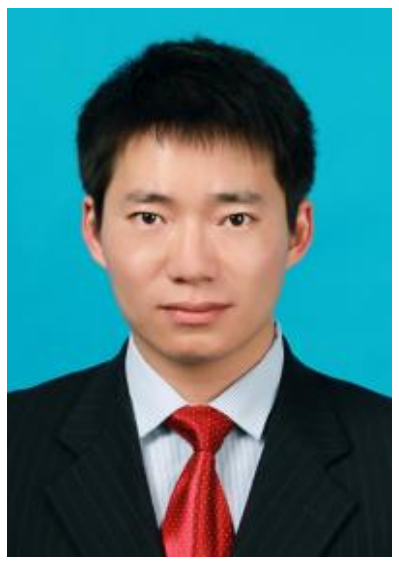

**Liwei Chen** received his M.S degree form Wuhan University, Wuhan, China, and the Ph.D. degree in mechanical system engineering from Tohoku University, Sendai, Japan, in 2019 and 2022, respectively.

He is currently a Postdoctoral Fellow at the University of Tokyo, Japan. His research interests include image processing and laser manufacturing.